\documentclass[journal]{IEEEtran}
\usepackage{float}
\usepackage{graphicx}
\usepackage{subfigure}
\usepackage{color}
\usepackage{amssymb}
\usepackage{amsmath}
\usepackage{algorithm}
\usepackage{algorithmic}
\usepackage{bm}
\usepackage[colorlinks]{hyperref}

\begin{document}
\title{Fluid Antenna-Assisted Simultaneous Wireless Information and Power Transfer Systems}
\author{Liaoshi Zhou, Junteng Yao, Tuo Wu, Ming Jin, Chau Yuen, \emph{Fellow, IEEE}, \\ and Fumiyuki Adachi, \emph{Life Fellow}, \emph{IEEE}

\thanks{L. Zhou, J. Yao, and M. Jin are with the faculty of Electrical Engineering and Computer Science, Ningbo University, Ningbo 315211, China (E-mail: \{2311100202, yaojunteng, jinming\}@nbu.edu.cn).}

\thanks{T. Wu and C. Yuen are with the School of Electrical and Electronic Engineering, Nanyang Technological University, 639798, Singapore (E-mail: $\rm \{tuo.wu, chau.yuen\}@ntu.edu.sg$).}

\thanks{F. Adachi is with the International Research Institute of Disaster Science (IRIDeS), Tohoku University, Sendai, Japan (E-mail: $\rm  adachi@ecei.tohoku.ac.jp$).}

}
\maketitle
\begin{abstract}
This paper examines a fluid antenna (FA)-assisted simultaneous wireless information and power transfer (SWIPT) system. Unlike traditional SWIPT systems with fixed-position antennas (FPAs), our FA-assisted system enables dynamic reconfiguration of the radio propagation environment by adjusting the positions of FAs. This capability enhances both energy harvesting and communication performance. The system comprises a base station (BS) equipped with multiple FAs that transmit signals to an energy receiver (ER) and an information receiver (IR), both equipped with a single FA. Our objective is to maximize the communication rate between the BS and the IR while satisfying the harvested power requirement of the ER. This involves jointly optimizing the BS's transmit beamforming and the positions of all FAs. To address this complex convex optimization problem, we employ an alternating optimization (AO) approach, decomposing it into three sub-problems and solving them iteratively using first and second-order Taylor expansions. Simulation results validate the effectiveness of our proposed FA-assisted SWIPT system, demonstrating significant performance improvements over traditional FPA-based systems.
\end{abstract}
\section{Introduction}
\IEEEPARstart{R}{ecently}, fluid antenna (FA)-assisted wireless communication has garnered  substantial research attention, offering the ability to dynamically adjust antenna positions and thus achieve higher data transmission rates for next-generation communication systems \cite{WMa23,LZhu23,JYao24,KKWong22,KKWong23}. Unlike traditional systems that utilize fixed-position antennas (FPAs) and are limited in fully exploiting spatial resources due to their immobility, FA-assisted systems represent a new paradigm. These systems can actively reshape the wireless propagation channel by maneuvering antennas within a designated area, providing increased spatial degrees of freedom (DoFs) and achieving enhanced communication rates \cite{LZhu24,XLai24}. Consequently, FAs have been integrated into various advanced wireless systems, such as FA-assisted multi-user systems \cite{HQin24}, FA-assisted over-the-air computation (AirComp) systems \cite{DZhang24}, and FA-assisted non-orthogonal multiple access (NOMA) systems \cite{JZheng24}.

On the other hand, wireless communication networks are increasingly focused on achieving energy-efficient communication.  Information transmission enabled simultaneous wireless information and power transfer (SWIPT) has emerged as a promising energy harvesting (EH) technique to address the energy scarcity problem in future energy-hungry wireless communication systems \cite{CPan20,CHu22}. The fundamental concept of SWIPT involves a transmitter with a dedicated power supply sending wireless signals to information receivers (IRs) and energy receivers (ERs). While IRs decode the information, ERs harvest energy from these signals. Given its advantages, SWIPT has been extensively explored across various contexts, including  NOMA  systems \cite{JRen23}, cognitive radio systems \cite{CHu22}, and secure systems \cite{ZZhu23}.

Inspired by these benefits, integrating  FA into SWIPT systems naturally enhances communication rates and energy efficiency within wireless networks. Unlike traditional SWIPT systems equipped with FPAs, where channel characteristics between the IRs/ERs are static and performance gains are limited to transmit beamforming, FAs offer the possibility to actively reshape these channels. By strategically optimizing the positions of FAs at both the transmitter and the receivers, improvements can be achieved in signal power, harvested power, and beamforming flexibility. To the best of our knowledge, a comprehensive optimization investigation into FA-assisted SWIPT systems has not yet been conducted.

To address this research gap, we introduce a FA-assisted SWIPT system, featuring a base station (BS) equipped with multiple FAs transmitting signals to an IR with a single FA and an ER with a single FA. Our objective is to maximize the IR's communication rate, considering the harvested power requirement of the ER and the power constraints of the BS. This involves jointly optimizing the BS's transmit beamforming, the positions of the BS's transmit FAs, and the positions of the IR's and ER's receive FAs. To tackle this highly non-convex problem, we employ an alternating optimization (AO) algorithm, utilizing first-order and second-order Taylor expansions to precisely adjust the positions of all FAs. Simulation results confirm the significant performance benefits of our proposed scheme over benchmarks, particularly in terms of energy harvesting and communication efficacy.

\textit{Notations}: $\mathrm{Tr}(\mathbf{A})$, $\mathbf{A}^H$, and $\mathbf{A}^T$ denote the spectral norm, trace, conjugate transpose, and transpose, respectively; $\left\|\mathbf{a}\right\|_2$ denotes the 2-norm of vector $\mathbf{a}$; $\mathbf{A}\succeq \mathbf{0}$ indicates that $\mathbf{A}$ is positive semidefinite; $\Re\{x\}$ means the real part of $x$; $\mathcal{CN}(0, \sigma^2)$ denotes the distribution of a circularly symmetric complex Gaussian variable with mean $0$ and variance $\sigma^2$.

\section{ System Model and Problem Formulation}

\begin{figure}[h]
\centering
\includegraphics[width=2.5in]{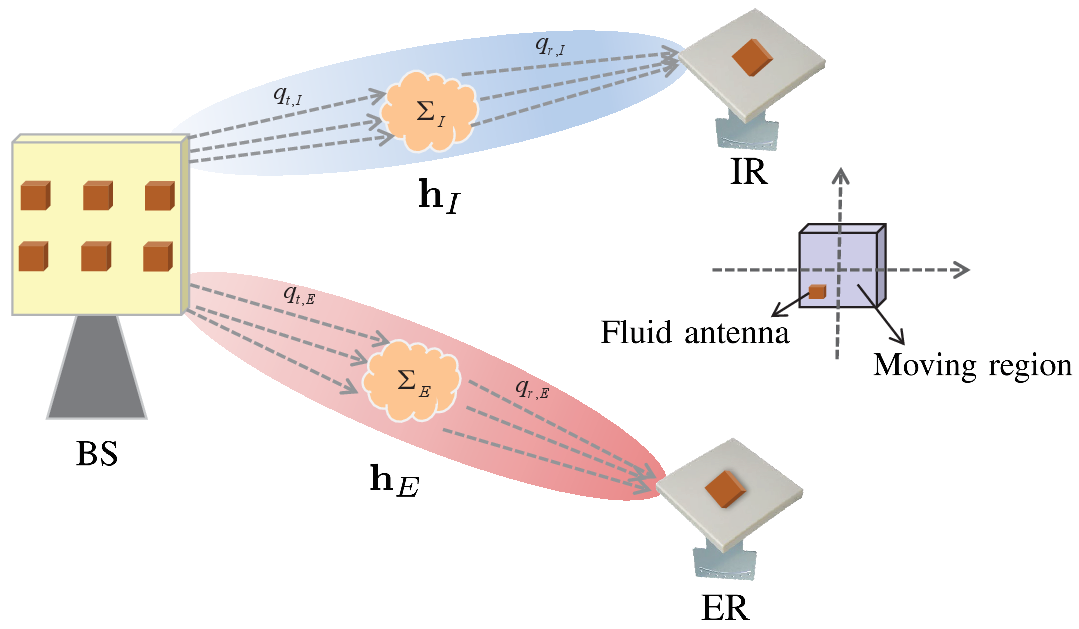}
\caption {The system model of FA-assisted SWIPT.}
\label{model1}
\end{figure}

As shown in Fig.~\ref{model1}, we consider a FA-assisted SWIPT system, which contains a BS, an IR, and an ER. We assume that the BS is equipped with $M (M\geq 2)$ FAs, which can move freely within a certain range denoted as $\zeta_t$.{The IR and ER are both equipped with a single FA with corresponding certain movable ranges, which are represented as $\zeta_{r,I}$ and $\zeta_{r,E}$, respectively.} We introduce a two-dimensional Cartesian coordinate system to describe the locations of the FAs. Specifically, we use $\mathbf{t}_{m}=[x_m^t, y_m^t]^T, m\in\mathcal{M}=\{1,\cdots,M\}$ to denote the location of the $m$-th FA at the BS. Similarly, we use $\mathbf {r}_{i}=[x^r_{i},y^r_{i}]^T, i\in\{I,E\}$ to denote the location of the FA at the IR or the ER.

During the downlink transmission, the signal received  at the IR and the ER are given by
\begin{equation}\label{001}
\mathbf{y}_i = \mathbf{h}_{i}\mathbf{w}x+n_i, i\in\{I,E\}.
\end{equation}
Without loss of generality, we assume the data streams $x \sim\mathcal{CN}(0,1)$ and $\mathbf{w}\in \mathbb{C}^{M \times 1}$ is the transmit beamforming vector at the BS. $\mathbf{h}_{I}\in \mathbb{C}^{1 \times M}$ and $\mathbf{h}_{E}\in \mathbb{C}^{1 \times M}$ are the channels from the BS to the IR and the ER, respectively. $n_I \sim\mathcal{CN}(0,\sigma^{2}_{I})$ and $n_E \sim\mathcal{CN}(0,\sigma^{2}_{E})$ are the additive Gaussian noise at the IR and the ER, respectively.

In this paper, we consider a far-field wireless channel model where the size of the transmit/receive movable ranges are significantly smaller than the signal propagation distance. Consequently, we assume a planar wavefront, and the angles of arrival (AoA) and angles of departure (AoD) remain constant for each channel path component \cite{WMa23, LZhu23}. We denote the number of transmit paths for the BS-IR and BS-ER links as $q_{t,I}$ and $q_{t,E}$, respectively. For the transmit FAs of the BS, the signal propagation distance difference from the origin $O^t$ on its $k$-th transmit path ($k \in {1, \cdots, q_{t,i}}$) is defined as $\rho_{t,i}^k(\mathbf{t}) = x_{t}\sin\phi_{t,i}^k \cos\psi_{t,i}^k + y_t \cos\phi_{t,i}^k$, where $i \in \{I, E\}$. Here, $\phi_{t,I}^k$ and $\psi_{t,I}^k$ denote the elevation and azimuth AoDs of the $k$-th transmit path from the BS to the IR, while $\phi_{t,E}^k$ and $\psi_{t,E}^k$ represent those from the BS to the ER. Accordingly, the BS transmit field response vectors are then expressed as follows \cite{YYe24}
\begin{equation}\label{003}
\bm{\xi}_i\mathbf{(t)}= \left [e^{j \frac{2\pi}{\lambda}\rho_{t,i}^1(\mathbf{t})}, \cdots, e^{j \frac{2\pi}{\lambda}\rho_{t,i}^{q_{t,i}}(\mathbf{t})}\right]^T, i\in\{I,E\},
\end{equation}
where $\lambda$ is the carrier wavelength. Thus, the transmission field response matrices from the BS to the IR and the ER are given by
\begin{equation}\label{004}
\bm{\Xi}_i\mathbf{(\overline{t})} = \left [\bm{\xi}_i(\mathbf{t}_1),\bm{\xi}_i(\mathbf{t}_2), \cdots, \bm{\xi}_i(\mathbf{t}_M)\right], i\in\{I,E\}.
\end{equation}

Similarly, let us denote ${q_{r,I}}$ and ${q_{r,E}}$ as the number of receive paths for the BS-IR and BS-ER links, respectively. For the receive fluid antennas (FAs) of the IR and ER, the signal propagation difference from their respective reference points $O^r_i$ for the $s$-th receive path $(s\in\{1,\cdots, {q_{r,i}}\})$ is given by $\rho_{r,i}^s(\mathbf{r}_i)=x_{r,i} \sin\phi_{r,i}^s \cos\psi_{r,i}^s+y_{r,i} \cos\phi_{r,i}^s$, where
 $i \in \{I, E\}$. Here, $\phi_{r,i}^s$ and $\psi_{r,i}^s$ denote the receive elevation and azimuth AoAs of the $s$-th
path of the BS-IR and BS-ER links. Consequently, the receive field response vectors at the IR and ER are 
expressed as 
\begin{align}\label{005}
\mathbf{f}(\mathbf{r}_i) = \left [e^{j \frac{2\pi}{\lambda}\rho_{r,i}^1(\mathbf{r}{_i})},\cdots,e^{j \frac{2\pi}{\lambda}\rho_{r,i}^{q_{r,i}}(\mathbf{r}{_i})}\right]^T, i\in\{I,E\}.
\end{align} 

Moreover, let us define the path response matrix $\bm{\Sigma}_i\in \mathbb{C}^{q_{r, i} \times q_t}, i\in\{I,E\}$ as the response coefficients of all paths from the origin $O^t$ of the BS to the $O^r_i, i\in\{I,E\}$ of the IR/ER. Then, the channel vectors of the BS-IR and BS-ER links are given as
\begin{align}
\label{eq6a}\mathbf{h}_I=&\mathbf{f}^H(\mathbf{r}_I)\bm{\Sigma}_I\bm{\Xi}_I\mathbf{(\overline{t})},\\
\label{eq6b}\mathbf{h}_E=&\mathbf{f}^H(\mathbf{r}_E)\bm{\Sigma}_E\bm{\Xi}_E\mathbf{(\overline{t})},
\end{align}
respectively \footnote{ {The information of $\mathbf{h}_I$ and $\mathbf{h}_E$ can be determined by using some channel estimation methods, e.g., compressed sensing \cite{WMa231}.}}.

Building upon \eqref{eq6a}, the communication rate at the IR is
\begin{equation}
R=\log_2\left(1+\frac{\mathbf{h}_I\mathbf{W}\mathbf{h}^H_I}{\sigma^{2}_{I}}\right),
\end{equation}
where $\mathbf{W}=\mathbf{ww}^H$ is rank-one matrix.

For the ER, we consider a linear energy harvesting model, and the total harvested power at the ER is given by \cite{CPan20}, \cite{ZZhu23}
\begin{equation}\label{007}
Q=\tau\mathrm{Tr}\left(\mathbf{h}_E \mathbf{W} \mathbf{h}_E^H\right),
\end{equation}
where $0<\tau\leq1$ is the energy harvesting efficiency.

In this paper, we aim to maximize the communication rate between the BS and IR under the requirement of the harvested power of the ER by jointly optimizing the transmit beamforming of the BS, the positions of FAs of the BS, the IR, and the ER. The corresponding optimization problem can be expressed as
\begin{subequations}\label{008}
\begin{align}
\max\limits_{\mathbf{\overline{t}},\mathbf{r}_i,\mathbf{W}\succeq\mathbf{0}} \quad &R \label{02a}\\
\mathrm{s.t.} \quad \ &\mathbf{\overline{t}} \in \zeta_t, \label{02b}\\
&\mathbf{r}_i \in \zeta_{r,i}, i\in\{I,E\}, \label{02c}\\
&||\mathbf{t}_m-\mathbf{t}_v||_2\geq D,~m,v\in\mathcal{M},~m\neq v, \label{02d}\\
&\mathrm{Tr}(\mathbf{W}) \leq P_{\max}, \label{02e}\\
&Q\geq\overline{Q}, \label{02f}\\
&\mathrm{rank}(\mathbf{W})=1, \label{02g}
\end{align}
\end{subequations}
where \eqref{02b} and \eqref{02c} are the movable range constraints of the FAs of the BS, the IR, and the ER; \eqref{02d} represents the minimum distance constraint for the transmitting FAs to prevent coupling; \eqref{02e} denotes the maximum transmit power constraint at the BS. \eqref{02f} means the power received at the ER should be larger than a certain value $\overline{Q}$, where $\overline{Q}$ is the harvested power threshold at the ER. Since the optimization variables are coupled in \eqref{02a}, \eqref{02d} and \eqref{02f}, Problem \eqref{008} is non-convex, which is difficult to obtain an optimal solution.
\section{Proposed Algorithm}
 {In this section, we decompose Problem \eqref{008} into three sub-problems about transmit beamforming,  transmit FAs' locations, and receive FA's locations, and use the AO algorithm to  iteratively optimize these sub-problems to obtain the locally optimal solution.}
\subsection{Optimization of Transmit Beamforming of BS}
Given $\mathbf{\overline{t}}$, $\mathbf{r}_I$, and $\mathbf{r}_E$, Problem \eqref{008} is non-convex due to the rank-one constraint \eqref{02g}. By omitting the constraint  \eqref{02g}, Problem \eqref{008} can be rewritten as
\begin{subequations}\label{009}
\begin{align}
\max\limits_{\mathbf{W}\succeq\mathbf{0}} \quad &R \label{009a}\\
\mathrm{s.t.} \quad \ &\eqref{02e}, \eqref{02f}.
\end{align}
\end{subequations}
It is readily to find that the objective function is concave with respect to $\mathbf{W}$, and the constraints \eqref{02e} and \eqref{02f} are both linear. Thus, Problem \eqref{009} is convex, which can be solved by the CVX toolbox \cite{MGrant}. After obtaining the optimal $\mathbf{W}$, we can reconstruct the rank-one solution by using Gaussian randomization \cite{BWei24}.

\subsection{Optimization of Transmit FAs' Locations of BS}
Given $\mathbf{W}$, $\mathbf{r}_I$, and $\mathbf{r}_E$, our aim is to optimize the position of the transmit FAs $\mathbf{\overline{t}}$. Since $\log_2(1 + x)$ is an increasing function with respect to $x$, maximizing $R$ is equivalent to maximizing $\mathrm{Tr}\left(\mathbf{h}_I\mathbf{W}\mathbf{h}^H_I\right)$. Thus, Problem \eqref{008} can be reformulated as
\begin{subequations}\label{010}
\begin{align}
\max\limits_{\mathbf{\overline{t}}} \quad& \mathrm{Tr}\left(\mathbf{h}_I\mathbf{W}\mathbf{h}^H_I\right) \label{010a}\\
\mathrm{s.t.} \quad  &~~~ \eqref{02b},\eqref{02d},\eqref{02f} \label{010b}.
\end{align}
\end{subequations}
We can see that the objective function $\mathrm{Tr}\left(\mathbf{h}_I\mathbf{W}\mathbf{h}^H_I\right)$ and $\mathrm{Tr}\left(\mathbf{h}_E \mathbf{W} \mathbf{h}_E^H\right)$ in \eqref{02f} have the same form. Therefore, we can handle them by using the same approach. Specifically, utilizing the property of the trace of a matrix, we can rewrite them as
\begin{align}\label{011}
\mathrm{Tr}\left(\mathbf{h}_i^H\mathbf{h}_i\mathbf{W}\right)=&\mathrm{Tr}\left(\sum_{m=1}^{M}{\gamma}_i(\mathbf{t}_m){\gamma}_i^H(\mathbf{t}_m)\mathbf{W} \right)\nonumber\\
=&\left(\eta_i+\gamma_i(\mathbf{t}_m)\gamma_i^H(\mathbf{t}_m)\right)\mathrm{Tr}\left(\mathbf{W} \right)\nonumber\\
=&\left(\eta_i+\underbrace{\bm{\xi}_i^H(\mathbf{t}_m)\bm{\omega}_i\bm{\xi}_i(\mathbf{t}_m)}_{G_i(\mathbf{t}_m)}\right)\mathrm{Tr}\left(\mathbf{W} \right),
 i\in\{I,E\},
\end{align}
where
\begin{align}
\label{013}\eta_i=&\sum_{j\neq m}^{M}{\gamma}_i(\mathbf{t}_j){\gamma}_i^H(\mathbf{t}_j), i\in\{I,E\},\\
\gamma_i(\mathbf{t}_m)=&\mathbf{f}^H(\mathbf{r}_i)\mathbf{\Sigma}_i\bm{\xi}_i(\mathbf{t}_m),  i\in\{I,E\},\\
\bm{\omega}_i=&\bm{\Sigma}_i^H\mathbf{f}(\mathbf{r}_i)\mathbf{f}^H(\mathbf{r}_i)\bm{\Sigma}_i,  i\in\{I,E\},
\end{align}
Since $\{\mathbf{t}_j,j\neq m\}_{j=1}^{M}$ is known, $\eta_i$ in \eqref{013} is a scalar that is not related with $\mathbf{t}_m$. Therefore, we only need to focus on $G_i(\mathbf{t}_m)$. It can be observed that $G_i(\mathbf{t}_m)$ is a convex function of $\bm{\xi}_i(\mathbf{t}_m)$, we can obtain a lower bound for $G_i(\mathbf{t}_m)$ by employing a first-order Taylor expansion at point $\widetilde{\mathbf{t}}_m$, which is given by
\begin{align}\label{015}
G_i(\mathbf{t}_m;\widetilde{\mathbf{t}}_m)=2\underbrace{\Re\left\{\bm{\xi}_i^H(\widetilde{\mathbf{t}}_m)\bm{\omega}_i\bm{\xi}_i(\mathbf{t}_m)\right\}}_{\widehat{G}_i(\mathbf{t}_m)}-\mu_i, i\in\{I,E\},
\end{align}
where $\mu_i=\bm{\xi}_i^H(\widetilde{\mathbf{t}}_m)\bm{\omega}_i\bm{\xi}_i(\widetilde{\mathbf{t}}_m)$. Since $\widehat{G}_i(\mathbf{t}_m)$ is neither convex nor concave with respect to $\mathbf{t}_m$, we utilize a second-order Taylor expansion to construct a surrogate function for $\widehat{G}_i(\mathbf{t}_m), i\in\{I,E\}$ \cite{LZhu23}. Then, we obtain the following equation \cite{WMa23}
\begin{align}\label{016}
\widehat{G}_i(\mathbf{t}_m;\widetilde{\mathbf{t}}_m) = &\widehat{G}_i(\widetilde{\mathbf{t}}_m)+\nabla\widehat{G}_i(\widetilde{\mathbf{t}}_m)^T(\mathbf{t}_m-\widetilde{\mathbf{t}}_m)\nonumber\\
&-\frac{{\kappa}_m^i}{2}(\mathbf{t}_m-\widetilde{\mathbf{t}}_m)^T(\mathbf{t}_m-\widetilde{\mathbf{t}}_m),
\end{align}
where

\begin{align}
\nabla\widehat{G}_i(\widetilde{\mathbf{t}}_m)=&\left[\frac{\partial\widehat{G}_i(\widetilde{\mathbf{t}}_m)}{\partial x^{t}_m},\frac{\partial\widehat{G}_i(\widetilde{\mathbf{t}}_m)}{\partial y^{t}_m}     \right],\\
\kappa_m^i=&\frac{8\pi^2}{\lambda^2}\sum_{k=1}^{q_{t,i}} \lvert \varrho^k_i \rvert,\\
\frac{\partial\widehat{G}_i(\widetilde{\mathbf{t}}_m)}{\partial x^{t}_m} =& - \frac{2\pi}{\lambda}\sum_{k=1}^{q_{t,i}} \lvert \varrho_i^k \rvert\sin\phi_{t,i}^k\cos\varphi_{t,i}^k\sin(\varsigma^k_i(\mathbf{\widetilde{t}}_m)), \\
\frac{\partial\widehat{G}_i(\widetilde{\mathbf{t}}_m)}{\partial y^{t}_m} =& - \frac{2\pi}{\lambda}\sum_{k=1}^{q_{t,i}} \lvert \varrho_i^k \rvert \cos\phi_{t,i}^k\sin(\varsigma^k_i(\mathbf{\widetilde{t}}_m)),\\
\varsigma^k_i(\mathbf{\widetilde{t}}_m)= &2\pi\rho_{t,i}^k(\mathbf{\widetilde{t}}_m)/\lambda-\angle\varrho_i^k,
\end{align}
$\lvert \varrho_i^k \rvert$ is amplitude for the $k$-th entry of $\bm{\xi}_i^H(\widetilde{\mathbf{t}}_m)\bm{\omega}_i$, and $\angle\varrho_i^k$ is the phase.

For the left side of constraint \eqref{02d}, we can relax it using a first-order Taylor expansion at point $\widetilde{\mathbf{t}}_m$ as the following form
\begin{align}\label{0017}
\beta(\mathbf{t}_m;\widetilde{\mathbf{t}}_m)=\frac{1}{||\widetilde{\mathbf{t}}_m-\mathbf{t}_v||_2}(\widetilde{\mathbf{t}}_m-\mathbf{t}_v)^T(\mathbf{t}_m-\mathbf{t}_v).
\end{align}

By removing the terms that are not related with $\mathbf{t}_m$, we can reformulate Problem \eqref{010} as
\begin{subequations}\label{017}
\begin{align}
\max\limits_{\mathbf{t}_m}&~~\widehat{G}_I(\mathbf{t}_m;\mathbf{t}_m^{(q)}) \label{017a}\\
\mathrm{s.t.} &~~\mathbf{\overline{t}} \in \zeta_t, \label{017b} \\
&~~\beta(\mathbf{t}_m;\mathbf{t}_m^{(q)})\geq{D},m,v\in\mathcal{M},~m\neq v,\label{017c}\\
&~~\widehat{G}_E(\mathbf{t}_m;\mathbf{t}_m^{(q)}) \geq\frac{\overline{Q}}{\tau\mathrm{Tr}(\mathbf{W})}+\mu_E-\eta_E \label{017d},
\end{align}
\end{subequations}
where $\mathbf{t}_m^{(q)}$ is the optimal $\mathbf{t}_m$ of the $q$-th iteration. It can be seen that the objective function \eqref{017a} is a concave quadratic function of $\mathbf{t}_m$, constraints \eqref{017b} and \eqref{017c} are linear, and \eqref{017d} is a convex constraint. Therefore, Problem \eqref{017} is convex.

\subsection{Optimization of Receive FA's Locations of IR and ER}
Given $\mathbf{W}$ and $\mathbf{\overline{t}}$, our aim is to optimize the FA's location at the IR and the FA's location at the ER. Because the optimization variables $\mathbf{r}_I$ and $\mathbf{r}_E$ are decoupled in Problem \eqref{008}, Problem \eqref{008} can be decomposed into two optimization problems
\begin{subequations}\label{018}
\begin{align}
\max\limits_{\mathbf{r}_I} \quad& \mathrm{Tr}\left(\mathbf{h}_I\mathbf{W}\mathbf{h}_I^H\right) \label{018a}\\
\mathrm{s.t.} \quad & \mathbf{r}_I \in \zeta_{r,I},
\end{align}
\end{subequations}
and
\begin{subequations}\label{019}
\begin{align}
\max\limits_{\mathbf{r}_E} \quad& \mathrm{Tr}\left(\mathbf{h}_E\mathbf{W}\mathbf{h}_E^H\right) \label{019a}\\
\mathrm{s.t.} \quad & \mathbf{r}_E \in \zeta_{r,E}.
\end{align}
\end{subequations}

Denote the eigenvalue decomposition (EVD) of $\mathbf{W}$ by $\mathbf{W}=\mathbf{U}_W\mathbf{\Lambda}_W\mathbf{U}_W^H$, we can get $\mathbf{Z}(\mathbf{r}_i)=\mathbf{h}_i\mathbf{U}_W\mathbf{\Lambda}_W^{\frac{1}{2}}\in\mathbb{C}^{1 \times M}$. Furthermore, we let $\mathbf{z}(\mathbf{r}_i)=\mathbf{\Lambda}_W^{\frac{1}{2}}\mathbf{U}_W^H\bm{\xi}_i^H\mathbf{(\overline{t})}\bm{\Sigma}^H\mathbf{f}(\mathbf{r}_i)\in\mathbb{C}^{M \times 1}$.
Rewrite the objective functions in Problem \eqref{018} and Problem \eqref{019} as
\begin{align}
\mathrm{Tr}\left(\mathbf{h}_i\mathbf{W}\mathbf{h}_i^H\right)=&\mathrm{Tr}\left(\mathbf{Z}(\mathbf{r}_i)\mathbf{Z}^H(\mathbf{r}_i) \right)\nonumber\\
=&\mathbf{z}^H(\mathbf{r}_i)\mathbf{z}(\mathbf{r}_i)\nonumber\\
=&\mathbf{f}^H(\mathbf{r}_i)\bm{\delta}_i\mathbf{f}(\mathbf{r}_i),  i\in\{I,E\},
\end{align}
where $\bm{\delta}_i=\bm{\Sigma}_i\bm{\xi}_i\mathbf{(\overline{t})}\mathbf{U}_W\bm{\Lambda}_W^{\frac{1}{2}}\bm{\Lambda}_W^{\frac{1}{2}}\mathbf{U}_W^H\bm{\xi}_i^H\mathbf{(\overline{t})}\bm{\Sigma}_i^H$.
Therefore, we can rewrite Problem \eqref{018} and Problem \eqref{019} as follow
\begin{subequations}\label{020}
\begin{align}
\max\limits_{\mathbf{r}_i} \quad& \mathbf{f}^H(\mathbf{r}_i)\bm{\delta}_i\mathbf{f}(\mathbf{r}_i) \label{020a}\\
\mathrm{s.t.} \quad & \mathbf{r}_i \in \zeta_{r,i},  i\in\{I,E\}.
\end{align}
\end{subequations}
It can be seen that the objective function in \eqref{020a} is similar to $G_i(\mathbf{t}_m)$ in \eqref{011}. Therefore, we can also use the second-order Taylor expansion method to solve \eqref{020a}, hence we do not repeat it here.

The procedures for solving Problem \eqref{008} is summarized in Algorithm 1, where the optimal  $\mathbf{\overline{t}}$, $\mathbf{{r}}_I$, $\mathbf{{r}}_E$, and $\mathbf{W}$ in the $l$th iteration are denoted as  $\mathbf{\overline{t}}^{(l)}$, $\mathbf{{r}}_I^{(l)}$, $\mathbf{{r}}_E^{(l)}$, and $\mathbf{W}^{(l)}$, respectively.

\begin{algorithm}[h]
\caption{The Proposed Alternating Optimization Algorithm}
\begin{algorithmic}[1]
\STATE \textbf{Initialize:} $l=0$, $\mathbf{\overline{t}}^{(0)}$, $\mathbf{{r}}_I^{(0)}$, and $\mathbf{{r}}_E^{(0)}$;
\STATE \textbf{Repeat} \\
\STATE\quad $l:=l+1$;         \\
\STATE \quad Update $\mathbf{W}^{(l)}$ by solving Problem \eqref{009} ;\\
\STATE \quad \textbf{Repeat} \\
\STATE \quad \quad $q:=q+1$;         \\
\STATE \quad \quad \textbf{for} $m=1\rightarrow M$  \\
\STATE \quad \quad\quad Update $\mathbf{\overline{t}}_m^{(q)}$ by solving Problem \eqref{017} ;\\
\STATE \quad \quad\textbf{end }
\STATE \quad \textbf{Until:} Convergence.\\
\STATE\quad Update $\mathbf{\overline{t}}^{(l)}$;
\STATE \quad  Update $\mathbf{{r}}_I^{(l)}$ and $\mathbf{{r}}_E^{(l)}$ by solving Problem \eqref{020};\\
\STATE \textbf{Until:} Convergence.
\end{algorithmic}
\end{algorithm}

\section{Numerical Result}
In our simulation experiments, we assume that the energy harvesting efficiency $\tau$ is set to $0.5$. The number of transmit and receive paths is set as $q_{t,I}=q_{t,E}=q_{r,I}=q_{r,E}=3$, and the elevation and azimuth angles $\phi_{t,i}^k,\psi_{t,i}^k,\phi_{r,i}^s,\psi_{r,i}^s, i\in\{I,E\}$ are all independent and identically distributed variables randomly distributed in $[0,\pi]$. The minimum distance between two FAs is set to be $D=\lambda/2$ with $\lambda=1$ m \cite{YYe24}, and the movable region for FAs at the BS is $\zeta_t=\left[-{\frac{A}{2}},{\frac{A}{2}}  \right] \times \left[-{\frac{A}{2}},{\frac{A}{2}}  \right]$, while the movable regions for the FA at the ER and the IR are both $\zeta_{r,I}=\zeta_{r,E}=\left[-{\frac{A}{4}},{\frac{A}{4}}  \right] \times \left[-{\frac{A}{4}},{\frac{A}{4}}  \right]$, where $A=4\lambda$ \cite{WMa23}. {Furthermore, we use the geometric channel model to describe the channels of the BS-IR and BS-ER links \cite{LZhu23}}. The path response matrixs $\bm{\Sigma}_I$ and $\bm{\Sigma}_E$ are assumed to be diagonal with $\bm{\Sigma}_i[1,1]\sim \mathcal{CN}({0}, {\nu/(\nu+1)})$ and $\bm{\Sigma}_i[p,p]\sim \mathcal{CN}({0}, {1/(\nu+1)(q_{r,i}-1)})$ for $p=2,3,\cdots,q_{r,i}, i\in\{I,E\}$, where $\nu$ represents the ratio of the average power
of the line-of-sight (LoS) path to that of the non-line-of-sight (NLoS) path and $\nu=1$ is assumed 
for the simulation \cite{WMa23}. We assume that the noise powers of the IR and the ER are $\sigma_I^2=\sigma_E^2=1$, the maximum transmit power to noise ratio is $P_{\max}/\sigma_I^2$ = 5 dB, and the harvested power threshold to noise ratio is $\overline{Q}/\sigma_E^2$ = 0 dB.

\begin{figure}[h]
\centering
\includegraphics[width=2in]{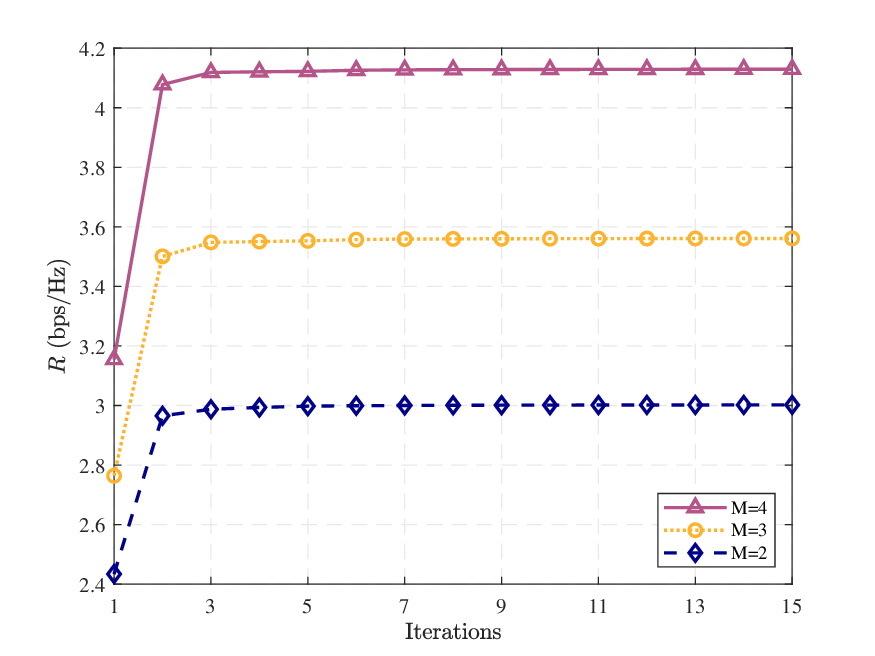}
\caption {Convergence of the proposed AO algorithm.}
\label{convergence}
\end{figure}

In Fig.~\ref{convergence}, we demonstrate the convergence behavior of our proposed  AO algorithm for $M = 2, 3, 4$. As observed from Fig.~\ref{convergence}, the AO algorithm converges after approximately 11 iterations. Additionally, the communication rate increases with the number of FAs, denoted by $M$. This increase is attributed to the greater DoFs provided by the FAs as $M$ increases, enhancing the system's capacity to optimize signal paths.

\begin{figure}[h]
\centering
\includegraphics[width=2in]{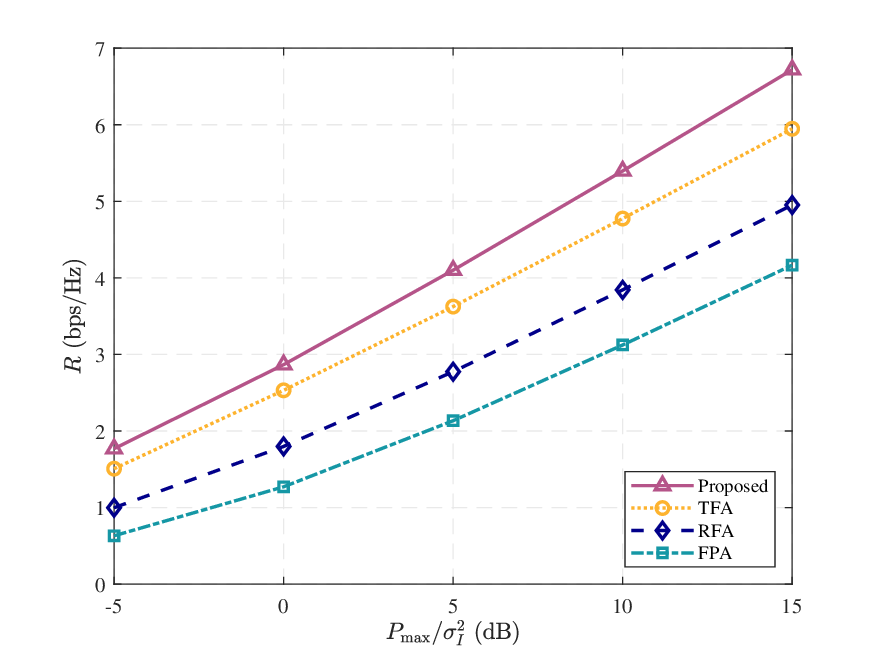}
\caption {$P_{\max}/\sigma^{2}_{I}$  versus $R$, where $M=4$.}
\label{SNR}
\end{figure}

\begin{figure}[h]
\centering
\includegraphics[width=2in]{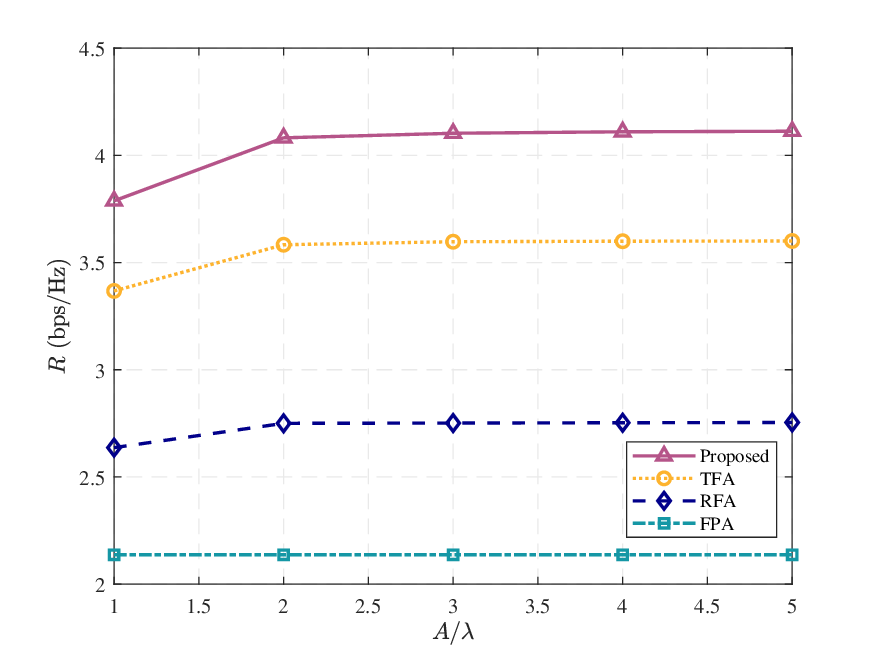}
\caption{$A/\lambda$ versus $R$, where $M=4$. }
\label{A}
\end{figure}

\begin{figure}[h]
\centering
\includegraphics[width=2in]{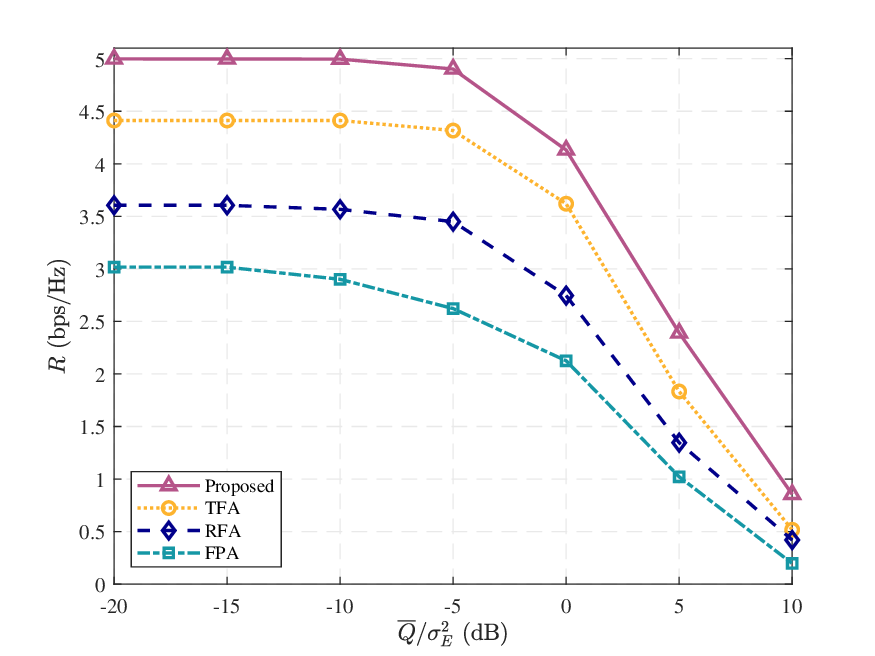}
\caption{$\overline{Q}/\sigma^{2}_{E}$ versus $R$, where $M=4$.}
\label{gamma}
\end{figure}

In Fig.~\ref{SNR}, Fig.~\ref{A}, and Fig.~\ref{gamma}, we compare the proposed scheme with the following benchmarks:

$\mathbf{transmit \ FA  \ (TFA)}$: The BS is equipped with $M$ FAs, while the IR and the ER are both equipped with a fixed single antenna.

$\mathbf{receive \ FA  \ (RFA)}$: The BS is equipped with $M$ fixed antennas, while the IR and the ER are both equipped with a single FA.

$\mathbf{FPA}$: The BS is equipped with $M$ fixed antennas, while the IR and the ER are bothe equipped with a single fixed antenna.

In Fig.~\ref{SNR}, we compare the communication rate $R$ of various schemes under different $P_{\max}/\sigma^{2}_{I}$ ratios, where $M=4$. It is observed that as the $P_{\max}/\sigma^{2}_{I}$ increases, the communication rate of all schemes increases. Additionally, our proposed scheme consistently outperforms the other schemes. Specifically, at $P_{\max}/\sigma^{2}_{I} = 10$ dB, our proposed scheme achieves improvements of $13.2\%$, $40.6\%$, and $73.1\%$ over the ``TFA'' scheme, the ``RFA'' scheme, and the ``FPA'' scheme, respectively.

In Fig.~\ref{A}, we investigate the impact of the normalized FAs' movable range $A/\lambda$ on the communication rate $R$, where $M=4$. The results indicate that, as $A/\lambda$ increases, the communication rates of all schemes except for the ``FPA'' scheme increase. This stagnation in the ``FPA'' scheme occurs because it employs fixed-position antennas (FPA). Notably, the communication rates of all schemes except for the ``FPA'' scheme converge when $A/\lambda$ exceeds 3. This suggests that the communication rate $R$ reaches its maximum within a limited movable range. Furthermore, our proposed scheme consistently outperforms the other benchmark schemes, demonstrating the advantages of utilizing FAs.

In Fig.~\ref{gamma}, we examine the impact of $\overline{Q}/\sigma^{2}_{E}$ on the communication rate $R$, where $M=4$. Observations from Fig.~\ref{gamma} reveal that as $\overline{Q}/\sigma^{2}_{E}$ increases, the communication rate $R$ decreases for all schemes. This decline is attributable to the higher energy harvesting requirement from the ER, which necessitates the BS to focus its transmission towards the ER, thus compromising the achievable rate $R$. Moreover, our proposed scheme demonstrates its capability to dynamically adjust the FAs' locations at the BS, the IR, and the ER to mitigate this impact, consistently achieving higher $R$ than other benchmark schemes under the same $\overline{Q}/\sigma^{2}_{E}$ conditions.

\section{Conclusion}
In this paper, we proposed a FA-assisted SWIPT system, where the BS, the IR, and the ER are both equipped with FA. By jointly designing the transmit beamforming of the BS and the positions of all FAs, we addressed the communication rate of the IR maximization problem under the constraint of the harvested energy of the ER. Due to the non-convex nature of the problem, we employ the AO algorithm to iteratively optimize the transmit beamforming and the positions of the FAs to obtain a sub-optimal solution. Numerical results show that the proposed FA-assisted SWIPT system achieves a higher communication rate compared to systems based on FPAs under the same harvested energy requirement of the ER.

\end{document}